\documentstyle[12pt]{article}
\begin{document}
\newcommand{\beq}{\begin{equation}}
\newcommand{\eeq}{\end{equation}}
\hoffset=-.8truecm
\title{Coulomb excitation at ultra-relativistic energies}
\author{B.F. Bayman\\
{\it School of Physics and Astronomy, University of Minnesota,}\\
{\it 116 Church Street S.E., Minneapolis, MN 55410, U.S.A.}\\
and\\
F.Zardi\\
{\it Istituto Nazionale di Fisica Nucleare and Dipartimento di Fisica,} \\
{\it Via Marzolo,8 I-35131 Padova,Italy.}}
\maketitle
\begin{abstract}
{\bf ABSTRACT} {\bf
In previous studies of the theory of Coulomb excitation, the term in the
electromagnetic interaction which is quadratic in the vector potential has been
ignored. In this paper, we use qualitative arguments and detailed numerical
calculations to show that this quadratic term must be included when the
bombarding energy is in excess of about 20 GeV per nucleon.} 
\end{abstract}

\section{Introduction}

The subject of relativistic Coulomb excitation has received extensive
study in the two decades since its theoretical foundations were established
in the classic work of Alder and Winther \cite{AW}. In recent years, 
this subject
has been shown to be relevant to the practical question of the stability of
beams of relativistic heavy ions, since some of the processes which lead to 
loss of beam ions are initiated by Coulomb excitation \cite{BRW,PRL,PSH}.

In the semi-quantal approach to relativistic nuclear Coulomb excitation, the
relative motion of the projectile and target is treated classically. Indeed, it
is usually assumed to be straight-line motion at constant speed $v$. The
evolution of the internal degrees of freedom of each nucleus, under the 
influence
of the classical electromagnetic field produced by the other nucleus, is then
followed using quantum mechanics. The effect of this classical
electromagnetic field on the nuclear charge and current densities has two
components: one that is linear in the electromagnetic potentials, and one that
is quadratic. All previous work on the theory of Coulomb excitation has been
based on the linear term. However, it is known that there are some
electromagnetic processes in which one cannot neglect the term quadratic in the
potentials. Examples are Thomson scattering of photons by electrons 
\cite{SAK}, and
the Zeeman effect in hydrogen atom states of high principal quantum number
\cite{GAS}. Our object in this paper is to determine whether the interaction 
term
quadratic in the potentials must be included when we study the nuclear effects
of the highly retarded electromagnetic field due to a rapidly passing
projectile.

\section{The electromagnetic interaction}

We follow the usual approach of the semi-classical theory of Coulomb
excitation, in which we solve the quantum-mechanical problem of the
response of the target nucleus to the perturbation provided by the
electromagnetic field of the passing projectile (ref.\cite{AW}). In the
absence of this perturbation, the target Hamiltonian has the form 
\begin{equation}
H_{0}=\sum_{j}~\frac{1}{2m}{\bf p}_{j}\cdot{\bf p}_{j}~+~W(\zeta)
\label{1.1}
\end{equation}
Here $\zeta$ represents the internal degrees of freedom of the target.
When the effect of the projectile electromagnetic field is included, the
target Hamiltonian becomes 
\begin{eqnarray}
H&=&\sum_{j}~\frac{1}{2m}({\bf p}_{j}-\frac{e_{j}}{c}{\bf A}^{{\rm
ret}}_{C}({\bf r}'_{j},t))^{2}+e_{j}\varphi^{{\rm
ret}}_{C}({\bf r}'_{j},t)+W(\zeta) \nonumber \\
~&=&H_{0}+V_{1}+V_{2}
\label{1.2} 
\end{eqnarray}
with
\begin{eqnarray}
V_{1}&=&-\sum_{j}\frac{e_{j}}{2mc}[~{\bf p}_{j}\cdot{\bf A}^{{\rm
ret}}_{C}({\bf r}'_{j},t)+{\bf A}^{{\rm
ret}}_{C}({\bf r}'_{j},t)\cdot{\bf p}_{j}~]\nonumber\\
&+&e_{j}\varphi^{{\rm ret}}_{C}({\bf r}'_{j},t),
\label{1.3}\\
V_{2}&=&\sum_{j}\frac{e_{j}^{2}}{2mc^{2}}{\bf A}^{{\rm
ret}}_{C}({\bf r}'_{j},t)\cdot{\bf A}^{{\rm
ret}}_{C}({\bf r}'_{j},t).
\label{1.4}
\end{eqnarray}
$e_{j}$ is the proton charge or 0 depending upon whether the jth nucleon is a
proton or neutron. Previous investigations (ref.\cite{WA,BZ}) of relativistic
Coulomb excitation have focussed only on the effect of the perturbation
$V_{1}$, the term linear in $e_{j}$. This term is evaluated using either the
Lienard-Wiechert potential for $\varphi^{{\rm ret}}_{C}({\bf r}'_{j},t)$ and 
${\bf A}^{{\rm ret}}_{C}({\bf r}'_{j},t)$, or the Fermi-Weissaker-Williams
method of equivalent photons \cite{JACK,BerBau}. In both approaches, 
it is assumed
that the perturbation produced by $V_{2}$, is of secondary
importance, and can be neglected.  We note that the classical non-relativistic
equations of motion, with the full Lorentz force,  
$$ m\frac{d {\bf v}}{d t}=e
{\bf E(r},t)+\frac{e}{c}{\bf v}\times{\bf B(r},t)~~.
$$
is obtained from the Hamiltonian (2.2) only if $V_{2}$ is included.

Our object in this study is to determine whether the $V_{2}$ term in
the interaction makes a numerically significant contribution to the Coulomb
excitation amplitude. We will show in the following that for lead
projectiles with kinetic energy below about 20 GeV per nucleon, the $V_{2}$ term
can safely be neglected. However, for kinetic energies at or above
80 GeV per nucleon, the contributions from $V_{2}$ are comparable to the
those from $V_{1}$. 

We assume that the perturbing electromagnetic field is produced by a
spherically-symmetric projectile of charge $Z_{P}e$ moving along a
trajectory given by 
\begin{equation}
{\bf r}={\bf b}~+~vt{\bf {\hat z}},
\label{1.5}
\end{equation} 
where ${\bf b}$ is the impact parameter vector perpendicular to ${\bf
{\hat z}}$, and $v$ is the constant projectile speed. The scalar and
vector potentials due to this projectile at the point ${\bf r'}$ of the
target at time $t$ can be taken to be the Lienard-Wiechert expressions
(ref.\cite{JACK}) 
\begin{eqnarray}
\varphi^{{\rm ret}}_{C}({\bf r'},t)&=&{Z_{_P}e\gamma \over 
\sqrt{(x-x')^{2}+(y-y')^{2}+\gamma^{2}(vt-z')^{2}}}\nonumber\\
&=&{Z_{_P}e\gamma \over 
\sqrt{|{\bf b}-\mbox{\boldmath$\rho$}'|^{2}+\gamma^{2}(vt-z')^{2}}},
\nonumber \\
{\bf A}^{{\rm
ret}}_{C}({\bf r'},t)&=&\frac{v}{c}~\varphi^{{\rm ret}}_{C}({\bf r'},t)
{\bf{\hat z}}.
\label{1.6}
\end{eqnarray}
We focus our attention on the Fourier transform of the matrix elements
of the perturbation, 
\begin{eqnarray}
V_{\beta \alpha}(\omega)&\equiv&\int^{\infty}_{-\infty}\frac{dt}{\hbar}e^{i
\omega t}<\phi_{\beta}|V_{1}(t)+V_{2}(t)|\phi_{\alpha}>\nonumber\\
&=&V^{(1)}_{\beta \alpha}(\omega)+V^{(2)}_{\beta \alpha}(\omega). 
\label{1.7}
\end{eqnarray}
Here $\phi_{\alpha}$ is an eigenstate of $H_{0}$ corresponding to
unperturbed eigenvalue $E_{\alpha}$. The first-order Born approximation
for the transition between $\phi_{\alpha}$ and $\phi_{\beta}$ is
expressed in terms this matrix element, with $\omega$ given its on-shell
value of $(E_{\beta}-E_{\alpha})/\hbar$. 

\section{Orders of magnitude}

We begin with preliminary comparisons of the relative magnitudes of
of matrix elements of $V_{1}$ and $V_{2}$ and of their dependences on
bombarding energy and $\omega$.  

{\it Bombarding energy dependence.} The Fourier transforms of $V_{1}$ and
$V_{2}$ require, respectively, the following time integrals:
$$
{\cal V}_{1}(|{\bf b}-\mbox{\boldmath$\rho$}'|,z')~\equiv~\int \frac{dt}
{\hbar}e^{i\omega t}~(\frac{\gamma}{\sqrt{|{\bf
b}-\mbox{\boldmath$\rho$}'|^{2}+\gamma^{2}|vt-z'|^{2}}}),
$$
$$
{\cal V}_{2}(|{\bf b}-\mbox{\boldmath$\rho$}'|,z')~\equiv~\int \frac{dt}
{\hbar}e^{i\omega t}~(\frac{\gamma^{2}}{|{\bf
b}-\mbox{\boldmath$\rho$}'|^{2}+\gamma^{2}|vt-z'|^{2}}).
$$

The two $t$ integrations yield
\begin{eqnarray}
{\cal V}_{1}(|{\bf
b}-\mbox{\boldmath$\rho$}'|,z')&=&\frac{e^{i\frac{\omega}{v}z'}}{\hbar
v}~K_{0}(\frac{\omega}{\gamma v}|{\bf b}-\mbox{\boldmath$\rho$}'|),\\
{\cal V}_{2}(|{\bf b}-\mbox{\boldmath$\rho$}'|,z')&=&\frac{e^{i\frac
{\omega}{v}z'}}
{\hbar v}~\pi~\gamma~\frac{e^{-\frac{\omega}{\gamma v}|{\bf b}
-\mbox{\boldmath$\rho$}'|}}{|{\bf b}-\mbox{\boldmath$\rho$}'|};
\end{eqnarray}
$V^{(1)}_{\beta \alpha}(\omega)$ and $V^{(2)}_{\beta \alpha}(\omega)$ are
obtained by calculating the matrix elements of these time integrals between the
nuclear states $\phi_{\alpha}$ and $\phi_{\beta}$.

$K_{0}(\frac{\omega}{\gamma v}|{\bf b}-\mbox{\boldmath$\rho$}'|)$ in Equation
(3.1) diverges like $\log (\gamma)$ as $\gamma \rightarrow \infty$, whereas
$\gamma~e^{-\frac{\omega}{\gamma v}|{\bf b}-\mbox{\boldmath$\rho$}'|}$ diverges
like $\gamma$. Thus as the bombarding energy becomes very large, the ${\bf A
\cdot A}$ matrix element grows relative to the matrix element linear in
($\varphi, {\bf A}$). Alternatively, we can say that the extra factor of 
$$
\frac{\gamma}{\sqrt{|{\bf
b}-\mbox{\boldmath$\rho$}'|^{2}+\gamma^{2}|vt-z'|^{2}}}
$$
in $V_{2}$ increases the effect of retardation, which is most dramatic at large
$\gamma$.

$\omega$-{\it dependence.}
For $\omega \rightarrow \infty$, both expressions (3.1) and (3.2) decay
exponentially. However, since $K_{0}(x) \rightarrow \sqrt{\frac{\pi}{2
x}}e^{-x}$, there is an extra factor of $1/\sqrt{\omega}$ in the
fall-off of $K_{0}(\frac{\omega}{\gamma v}|{\bf
b}-\mbox{\boldmath$\rho$}'|)$ compared to $e^{-\frac{\omega}{\gamma
v}|{\bf b}-\mbox{\boldmath$\rho$}'|}$. Thus at
large $\gamma$, where both matrix elements are appreciable, the ${\bf A
\cdot A}$ matrix element increases relative to the ($\varphi,
{\bf A}$) matrix element as $\omega$ increases. The extra retardation associated
with the ${\bf A \cdot A}$ term makes the electromagnetic pulse sharper, and so
favors higher $\omega$ values in the Fourier transform of the matrix element. 

{\it Relative magnitude.} We can get a crude estimate of the relative
importance of $V_2$ and $V_1$ by considering the ratio
$$
\frac{\frac{e^{2}}{2 m c^{2}} {\bf A}^{\rm ret} ({\bf r},t)\cdot {\bf A}^{\rm
ret} ({\bf r},t)}{e \varphi^{\rm ret}({\bf r},t)} \sim \frac{\frac{e^{2}}{2 m
c^{2}} (\frac{v}{c}\varphi^{\rm ret}({\bf r},t))^{2}}{e \varphi^{\rm ret}({\bf
r},t)}  
~\sim \frac{e}{2 m c^{2}}\varphi^{\rm ret}({\bf r},t)~(1-\frac{1}{\gamma^{2}}).
$$
To make this a little more quantitative, let us give $\varphi^{\rm ret}
({\bf r},t)$ its maximum value of $\gamma Z_{P}e/b$. Then the above ratio is
$$
\gamma Z_{P} \frac{e^{2}}{2 mc^{2} b}(1-\frac{1}{\gamma^{2}})
$$
For $Z_{P}=82$ and $b=10 Fm$, this
becomes $\sim .0063 \gamma (1-\frac{1}{\gamma^{2}})$. This has a value of $\sim
.01$ for $T_{P}/A=1$ GeV $(\gamma = 2.066)$, and $\sim .34$ for $T_{P}/A=50$ GeV
$ (\gamma =54.3)$. Thus we expect that $V_{1}$ can be neglected compared to
$V_{1}$ at projectile kinetic energies per nucleon of 1 GeV or lower, but at 50
GeV, $V_{2}$ may be of comparable importance to $V_{1}$. It will be seen below
that this expectation is verified by detailed calculations.

\section{Matrix elements}

The evaluation of $V^{(1)}_{\beta \alpha}(\omega)$ has been fully
described in the literature (ref.\cite{WA,BZ}). We therefore turn our
attention to $V^{(2)}_{\beta \alpha}(\omega)$: 
$$
V^{(2)}_{\beta \alpha}(\omega)~
=~\int^{\infty}_{-\infty} \frac{dt}{\hbar}e^{i\omega t}~
<\phi^{J_{\beta}}_{M_{\beta}}|\sum_{j}\frac{e_{j}^{2}}{2mc^{2}}
{\bf A}^{{\rm ret}}_{C}({\bf r}'_{j},t)\cdot{\bf A}^{{\rm
ret}}_{C}({\bf r}'_{j},t)|\phi^{J_{\alpha}}_{M_{\alpha}}> 
$$
$$
=~\frac{1}{2 \hbar m c^{2}}\int^{\infty}_{-\infty} dt e^ {i\omega
t}<\phi^{J_{\beta}}_{M_{\beta}}|\sum_{j}~e_{j}^{2}[~\varphi^{{\rm ret}}_
{C}({\bf r'}_{j},t)~]^{2}|\phi^{J_{\alpha}}_{M_{\alpha}}> 
$$
\beq
=~\frac{1}{2 \hbar m c^{2}}\int^{\infty}_{-\infty} dt e^
{i\omega t}[~\varphi^{{\rm ret}}_{C}({\bf
r'}_{j},t)~]~^{2}~\rho_{J_{\beta}M_{\beta};J_{\alpha}M_{\alpha}}({\bf r'}).
\label{1.8}
\eeq
The transition charge density introduced in the last line of Equation 
(\ref{1.8}) can be expanded in terms of spherical harmonics of 
${\bf{\hat r'}}$: 
\begin{eqnarray*}
\rho_{J_{\beta}M_{\beta};J_{\alpha}M_{\alpha}}({\bf r'})&\equiv&
<\phi^{J_{\beta}}_{M_{\beta}}|\sum_{j}e_{j}^{2} \delta ({\bf r'}-
{\bf r'}_{j})|\phi^{J_{\alpha}}_{M_{\alpha}}>\nonumber \\
&=&(-1)^{J_{\beta}-M_{\beta}}
\sum_{L}(J_{\beta}~J_{\alpha}~-M_{\beta}~M_{\alpha}
|L~M_{\alpha}-M_{\beta}~)
\rho_{L}(r')Y^{L}_{M_{\alpha}-M_{\beta}}({\bf {\hat r'}}),
\end{eqnarray*}
leading to a multipole expansion of our matrix element:
\begin{eqnarray}
V^{(2)}_{\beta \alpha}(\omega)&=&(-1)^{J_{\beta}-M_{\beta}}
\sum_{L}(J_{\beta}~J_{\alpha}~-M_{\beta}~M_{\alpha}|L~M_{\alpha}-M_{\beta}~)~
V^{L}_{M_{\alpha}-M_{\beta}}(\omega) \nonumber \\
V^{L}_{M}(\omega)&=&\frac{v^{2}}{2 \hbar m c^{4}}\int d^{3}r'
\rho_{L}(r')Y^{L}_{M}({\bf {\hat r'}})
~\int^{\infty}_{-\infty}dt e^{i \omega t}
\frac{(Z_{P}e\gamma)^{2}}{|{\bf b}-\mbox{\boldmath$\rho$}'|^{2}+\gamma^{2}
(vt-z')^2}\nonumber\\
&=&\frac{\pi v}{2 \hbar m c^{4}}(Z_{P}e)^{2}\gamma\int d^{3}r'
\rho_{L}(r')Y^{L}_{M}({\bf {\hat r'}})e^{i \frac{\omega}{v}z'}\frac
{e^{-\frac{\omega}{\gamma v}|{\bf b}-\mbox{\boldmath$\rho$}'|}}
{|{\bf b}-\mbox{\boldmath$\rho$}'|}.
\label{1.9}
\end{eqnarray}
We disentangle the ${\bf b}$ and $\mbox{\boldmath$\rho$}'$ 
dependence of $|{\bf b}-\mbox{\boldmath$\rho$}'|$ by using the expansion
$$
\frac{e^{-\frac{\omega}{\gamma v}|{\bf b}-\mbox{\boldmath$\rho$}'|}}
{|{\bf b}-\mbox{\boldmath$
\rho$'}|}~=~-\frac{\omega}{\gamma v}\sum_{\ell=0}^{\infty}h^{(1)}_{\ell}(i
\frac{\omega}{\gamma v}b)~j_{\ell}(i\frac{\omega}{\gamma
v}\rho') 
$$
\beq
\times\sum_{m=-\ell}^{\ell}\frac{(\ell-m-1)!!(\ell+m+1)!!}{(\ell-m)!!
(\ell+m)!!}~e^{im(\phi-\phi')}
\label{1.10}
\eeq
which is valid when $\rho'<b$. To complete the evaluation of the ${\bf
r}'$ integration in (\ref{1.9}), we express $\rho_{L}(r')Y^{L}_{M}({\bf
{\hat r'}})$ in cylindrical coordinates by expanding
$\rho_{L}(r')Y^{L}_{M} ({\bf {\hat r'}})$ in terms of 3-dimensional
harmonic oscillator eigenfunctions 
\begin{equation}
\rho_{L}(r')Y^{L}_{M}({\bf {\hat r'}})~=~\sum_{n}~c(n)~\psi^{nL}_{M}({\bf r}')
\label{1.11}
\end{equation}
and then performing a unitary transformation to products of 2-dimensional
oscillator eigenfunctions $\psi_{N_{\perp}M}(\rho',\varphi')$ and
one-dimensional oscillator eigenfunctions $\psi_{N_{z}}(z')$:
$$
\psi^{nL}_{M}({\bf r}')~=~\sum_{N_{\perp},N_{z}}~<n~L~M|N_{\perp}~N_{z}~M>
~\psi_{N_{\perp}M}(\rho',\phi')~\psi_{N_{z}}(z').
$$
The coefficients needed for this expansion are given in
ref.(\cite{BBB}). The $\rho',\phi',z'$ integrations can now be done in
(\ref{1.9}), with the final result 
$$
V^{L}_{M}(\omega)~=~i^L\sqrt{\frac{2}{\nu}}\frac{\pi^2}{mc^2}(\frac{Z_{_P}e^2}
{\hbar c})^2~\hbar\omega~
~\sum_{\ell=|M|,|M|+2,\dots}(2\ell+1)h^{(1)}_\ell
(i\frac{\omega}{\gamma v}b)
\frac{(\ell-M-1)!!}{(\ell-M)!!}\frac{(\ell+M-1)!!}{(\ell+M)!!}
$$
$$
\times\sum_{n=0}^{\kappa_{\rm max}}(-1)^nc(n)
\sum_{p=0,1,2\dots}^{n+\frac{L-|M|}{2}}(-1)^p{\tilde \psi}_{N_z}
(\frac{\omega}{v})\sqrt{p!(M+p)!}
<nLM|~2p+|M|,2n+L-|M|-2p,M>
$$
\begin{equation}
\sum_{j=0}^p \sum_{n'=0}^\infty~(-1)^j 
(\frac{\omega}{\gamma v\sqrt{\nu}})^{\ell+2n'}
~\frac 
{(M+\ell+2j+2n')!!}
{j!(M+j)!(p-j)!(2n')!!(2\ell+2n'+1)!!}~.
\label{1.12}
\end{equation}
In this equation we have set $\phi=\pi/2$ (the projectile trajectory is
in the ${\bf {\hat y}}-{\bf {\hat z}}$ plane), and assumed that both
$\omega$ and $M$ are positive. We get the remaining matrix elements
using the symmetry relations 
$$
V^{L}_{M}(-\omega)=(-1)^{L-M}V^{L}_{M}(\omega)~,~~~~~~~~~~~~
V^{L}_{-M}(\omega)=V^{L}_{M}(\omega);
$$
$\nu$ in (\ref{1.12}) is the harmonic oscillator size parameter
$\hbar/(m \omega_{osc})$, which is used for all the oscillator
eigenstates. It is chosen to give the best convergence for the expansion
(\ref{1.11}) of the transition charge density.  ${\tilde
\psi}_{N_z}(\frac{\omega}{v})$ is an oscillator eigenstate with argument
$\frac{\omega}{v}$, but with size parameter $1/\nu$. In the examples
given below, we will consider Coulomb excitation of a giant quadrupole
excitation in $^{40}$Ca (ref.\cite{BZ}). We use the Tassie model 
ref.\cite{TASS,SUZROW}, which describes the 2$^{+}$ resonance as a
one-quantum vibrational oscillation of an incompressible irrotational
fluid. Explicit expressions for the transition charge and current
densities are given in ref.\cite{BZ}. In this case, expansion of the
transition charge density in terms of harmonic oscillator eigenstates is
facilitated by the finite sum 
$$
(\sqrt{\nu}r')^{L+2 \kappa}e^{-\frac{\nu r'^{2}}{2}}Y^{L}_{M}({\bf {\hat r'}})
=\frac{\kappa!(2L+2
\kappa+1)!!}{2^{\kappa}}\sqrt{\frac{\sqrt{\pi}}{\nu^{\frac{3}{2}}}}
$$
$$
\times\sum_{n=0}^{\kappa}\frac{(-1)^{n}}{(\kappa-n)!}\frac{1}{\sqrt{n!(2
L +2n+1)!!2^{L+2-n}}}\psi^{nL}_{M}({\bf r}')  .
$$

\section{Application to the exitation of vibrational states}

The $V_{2}$ matrix elements calculated in Equation (4.5), added to the
well-known matrix elements of $V_{1}$, can be used in a full coupled-channel
calculation to yield a complete description of the Coulomb
excitation process. Since our object here is only to assess the relative
importance of $V_{2}$, we will limit our attention to the exactly-solvable
vibrational model. The main result of this model is that the probability,
$P(b,n)$, of exciting a state with $n$ oscillator quanta when the impact
parameter is $b$ is given by a Poisson distribution
$$ 
P(b,n)=e^{-q(b)}\frac{q(b)^{n}}{n!},
$$
where $q(b)$ is the square of the on-shell matrix element of the 
interaction between the
$n=0$ and $n=1$ states. The Coulomb excitation
cross-section for the population of the $n-$quantum state is then obtained from
$P(b,n)$ by an integration over $b$:
$$
\sigma_{n}=\int_{b_{{\rm min}}}^{\infty}2 \pi b db~P(b,n).
$$
The lower limit $b_{{\rm min}}$ is chosen to be large enough to ensure that only
electromagnetic interactions contribute to the process. We apply this model
to a hypothetical giant quadrupole oscillator band in $^{40}$Ca, with $\hbar
\omega$=20 MeV. The B(E2, $0^{+}\rightarrow 2^{+}$) is assumed to have a value
of 450 e$^{2}$fm$^{4}$, which exhausts the energy-weighted sum rule. The
projectile is $^{208}$Pb, and we have taken $b_{{\rm min}}$ to be 12
fm \cite{EML}. 

Figure 1 shows calculated excitation functions for populating
the magnetic substates of the one-quantum $I=2$ level, using $V_{1}$ alone.
The curves labelled by $M=|1|,|2|$ correspond to the summed cross-sections
for $M=\pm1,~\pm2$, respectively. The strong decrease of the $M=0$
cross-section at high bombarding energy is a result of cancellation of the
contributions of the scalar and vector potentials to $V_{1}$ of (2.3). At the
highest energies, the $M=\pm1$ cross-section is greatest. This is related to
the applicability of the equivalent photon method in this region, since a
shower of photons moving in the $+{\hat z}$ direction can only have $M=\pm1$. 

Figure 2 shows calculated excitation functions, including the effects of both
$V_{1}$ and $V_{2}$. Since the ${\bf A \cdot A}$ interaction essentially
involves the delivery of two photons to the target, arguments limited to
one-photon exchange no longer apply. In particular, the $M=0$ cross-section is
no longer suppressed at high bombarding energy. The large cross-section is due
to the fact that the strongly-retarded potential can produce $M=0$ transitions
at very large impact parameters. 

A polarization-insensitive measurement would yield the cross-section
incoherently summed over $M$. This is plotted in Figure 3, for $V_{1}$ alone,
and for the full $V_{1}+V_{2}$ interaction. It is seen that for bombarding
energy per nucleon below about 20 GeV, $V_{2}$ makes a relatively small
contribution, but at 80 GeV the full $V_{1}+V_{2}$ interaction yields
about twice the cross-section of $V_{1}$ alone. At higher bombarding energy,
the $V_{2}$ interaction dominates. 

\section{Discussion}

We have seen that qualitative arguments and detailed numerical calculations
strongly indicate the importance of  the ${\bf A \cdot A}$ contribution to the
response of a target nucleus to the electromagnetic field of a
highly-relativistic projectile. It should be emphasized that approximation
methods such as the FWW method of virtual quanta provide convenient
descriptions of the electromagnetic field, but do not, by themselves, include
the effect of ${\bf A \cdot A}$ which is quadratic in $e_j$.

An uncertainty of our analysis is the applicability of the non-
relativistic Hamiltonian (2.2) when the external electromagnetic field 
is very strong, even when the protons in the target are moving slowly
. If we attempt to derive this Hamiltonian by taking the non-
relativistic limit of a proton  Dirac equation, we need to assume that 
both the proton kinetic energy and $e\varphi^{{\rm ret}}_{C}$ are small 
compared to the proton rest energy (see, for example, \cite{BET}). But we 
have seen in Section III that the condition $e \varphi^{{\rm ret}}_{C}
<<mc^{2}$ is violated for grazing collisions when the bombarding energy per 
nucleon exceeds about 50 GeV. Thus, if nucleons are to be regarded as 
slowly-moving Dirac particles, the discussion of their electromagnetic 
properties requires something more complicated than Equation (2.2) when we 
are dealing with the fields encountered in ultra-relativisitic collisions.

The $V_{2}$ contributions to the curves in Figures 2 and 3 suffer from a gauge
ambiguity. The on-shell matrix elements of $V_{1}$ are gauge invariant; the
on-shell matrix elements of $V_{2}$ are not. Of course, if we use $V_{2}$ in a
full coupled-channel calculation, and arrive at exact Coulomb-excitation
cross-sections, these will be manifestly gauge invariant. We note that a full
coupled-channel calculation using $V_{1}$ alone will {\it not} be gauge
invariant, since both $V_{1}$ and $V_{2}$ are required in Equation (2.2) in
order to yield a gauge-invariant Schrodinger equation. The vibrational model
used in this paper is not fully gauge invariant. The gauge we have used is that
implied by the Lienard-Wiechert potential, and is widely employed in studies of
Coulomb excitation.  Although we cannot claim gauge invariance for our results,
we believe that we have shown that, at bombarding energies in 
excess of about 20
GeV per nucleon, it is important to include the effects of the 
${\bf A \cdot A}$
component of the electromagnetic interaction.

\vskip 1.5truecm
{\bf Figure captions}

\vskip .5truecm
{\bf Fig.1} Coulomb excitation cross sections for one quadrupole phonon for the
standard term $V_1$. The curves correspond to different values of magnetic
quantum number transfer. More details can be found in the text.
\vskip .2truecm
{\bf Fig.2} Coulomb excitation cross sections of one quadrupole phonon for the
full interaction $V_1+V_2$.
\vskip .2truecm
{\bf Fig.3} Comparison of the cross sections for $V_1$ and $V_2$ summed 
over the magnetic quantum numbers.

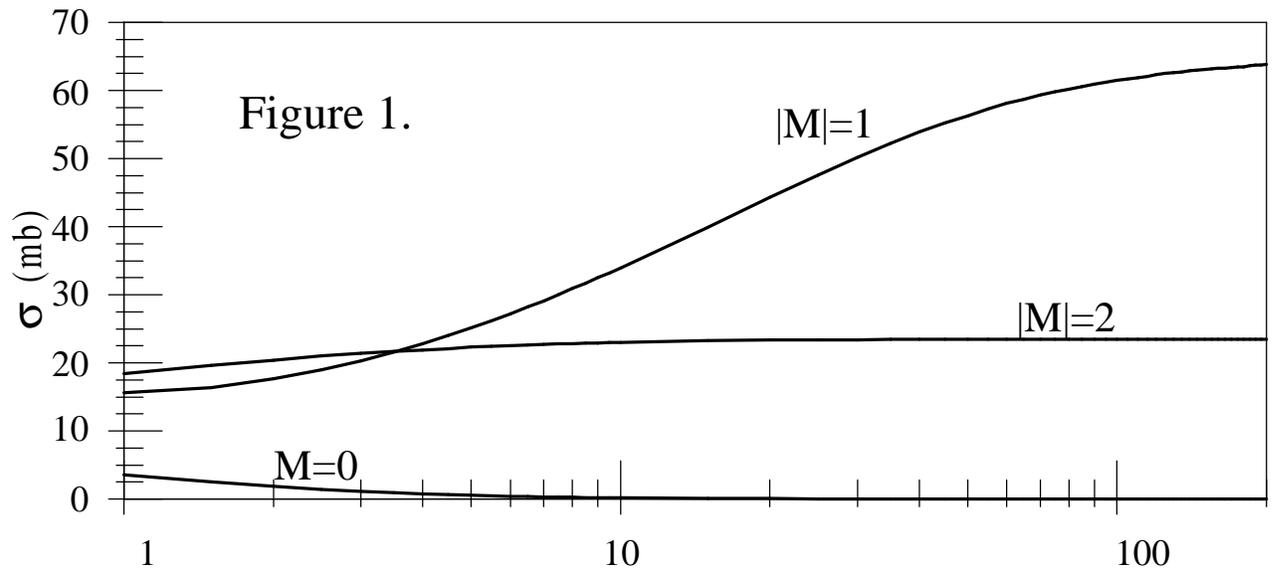
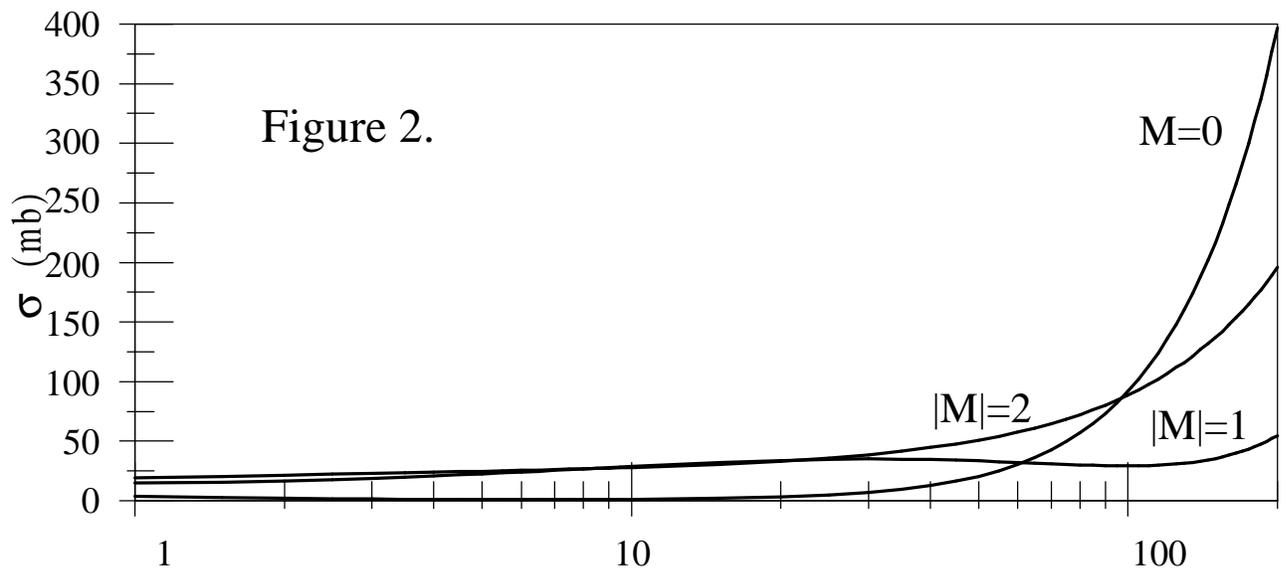
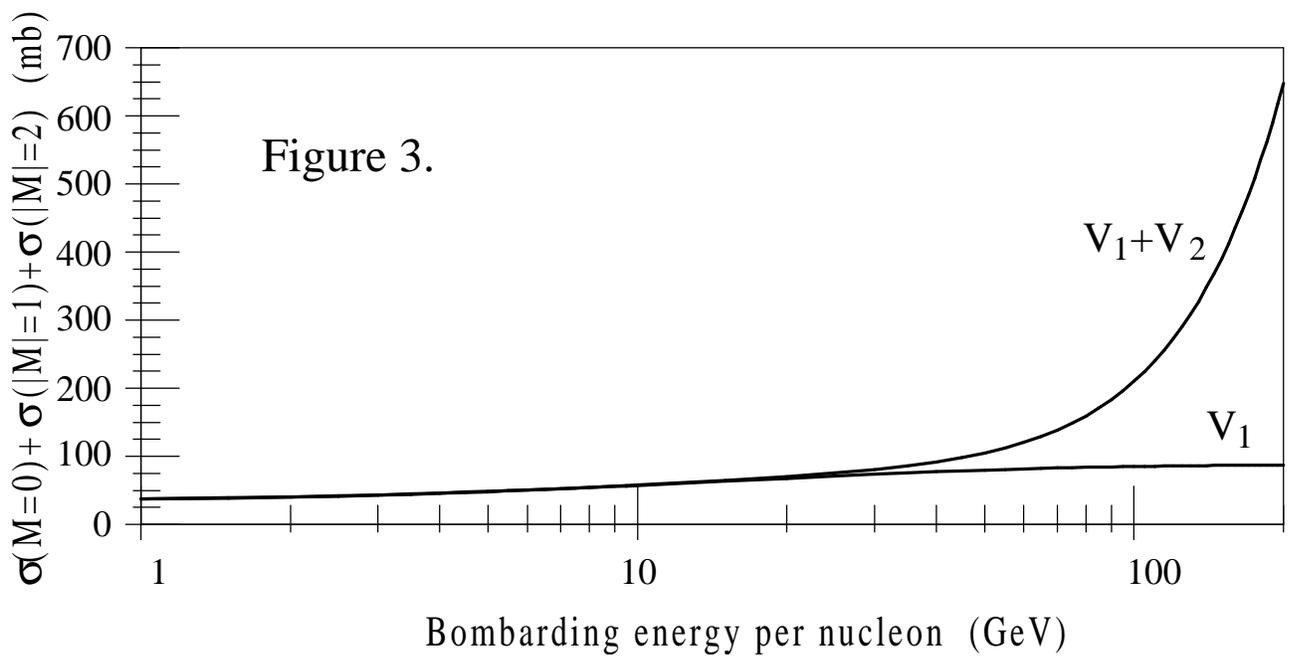
\end{document}